\documentclass[conference]{IEEEtran}
\usepackage{graphicx}
\usepackage{lipsum}
\usepackage{hyperref}
\usepackage{cite}
\usepackage{pifont}
\usepackage{enumitem}
\usepackage{caption}
\usepackage{subcaption}
\usepackage{array}
\newcolumntype{P}[1]{>{\centering\arraybackslash}p{#1}}
\usepackage{tabularx}

\IEEEoverridecommandlockouts
\usepackage{atbegshi}% http://ctan.org/pkg/atbegshi
\AtBeginDocument{\AtBeginShipoutNext{\AtBeginShipoutDiscard}}

\begin{document}

%+++++++++++++++++++++++++++++++++++++++++++
\title{\LARGE Energy-Efficient Hybrid Beamforming for Integrated Sensing and Communication Enabled mmWave MIMO Systems}
%\pagenumbering{arabic}
\author{
    \IEEEauthorblockN{Jitendra~Singh\IEEEauthorrefmark{1}, Suraj~Srivastava\IEEEauthorrefmark{2}, and Aditya~K.~Jagannatham\IEEEauthorrefmark{1}, }
    \IEEEauthorblockA{\IEEEauthorrefmark{1}Department of Electrical Engineering, Indian Institute of Technology Kanpur, India}
    
    \IEEEauthorblockA{\IEEEauthorrefmark{2}Department of Electrical Engineering, Indian Institute of Technology Jodhpur, India }
Email: \IEEEauthorrefmark{1}\{jitend, adityaj\}@iitk.ac.in \IEEEauthorrefmark{2}\{surajsri\}@iitj.ac.in \vspace{-5mm}}
%\thanks{%J. Singh and A. K. Jagannatham are with the Department of Electrical Engineering, Indian Institute of Technology Kanpur, Kanpur, UP 208016, India (e-mail: jitend@iitk.ac.in; adityaj@iitk.ac.in).}
%}
\thanks{%L. Hanzo is with the School of Electronics and Computer Science, University of Southampton, Southampton SO17 1BJ, U.K. (e-mail: lh@ecs.soton.ac.uk).}
}

\maketitle
\begin{abstract}
This paper conceives a hybrid beamforming (HBF) design that maximizes the energy efficiency (EE) of an integrated sensing and communication (ISAC)-enabled millimeter wave (mmWave) multiple-input multiple-output (MIMO) system. In the system under consideration, an ISAC base station (BS) with the hybrid MIMO architecture communicates with multiple users and simultaneously detects multiple targets. The proposed scheme seeks to maximize the EE of the system, considering the signal-to-interference and noise ratio (SINR) as the user's quality of service (QoS) and the sensing beampattern gain of the targets as constraints. To solve this non-convex problem, we initially adopt Dinkelbach’s method to convert the fractional objective function to subtractive form and subsequently obtain the sub-optimal fully-digital transmit beamformer by leveraging the principle of semi-definite relaxation. Subsequently, we propose a penalty-based manifold optimization scheme in conjunction with an alternating minimization method to determine the baseband (BB) and analog beamformers based on the designed fully-digital transmit beamformer.
Finally, simulation results are given to demonstrate the efficacy of our proposed algorithm with respect to the benchmarks.
\end{abstract}



\maketitle

\section{\uppercase{INTRODUCTION}}
\IEEEPARstart{H}{igh} data rate and accurate sensing are essential objectives for various applications of 6G wireless networks, such as vehicle-to-everything (V2X) communications, vehicle-to-infrastructure (V2I) networks, connected automated vehicles (CAVs), etc. Integrated sensing and communication (ISAC) with millimeter wave (mmWave) multi-input multi-output (MIMO) technology is a plausible solution to accomplish these objectives \cite{ISAC_1,ISAC_3}. By incorporating moderate hardware changes in the mmWave MIMO systems, one can develop an ISAC enabled mmWave MIMO system, which can yield the mutual benefits of sensing as well as communication. Furthermore, one can generate ultra-narrow beams by employing massive antenna elements at the ISAC base station (BS), which are highly desirable for precise sensing, and to also achieve the large array gain necessary to enhance the data rates at the users.

To shed light on these advantages, the authors of \cite{mm_ISAC_1,mm_ISAC_2,mm_ISAC_4,mm_ISAC_3,mm_ISAC_5} investigated various hybrid beamforming (HBF) designs for ISAC-aided mmWave MIMO systems to meet both the data rate and sensing requirements. In consonance with the HBF architecture, the signal processing burden is partitioned into baseband (BB) and analog/radio frequency (RF) domains, which significantly reduces the number of power-hungry RF chains (RFCs) required \cite{co_2,mmWave_MIMO_RIS}. Note that the analog domain beamformer in HBF is implemented using a digitally-controlled network of phase shifters.
To maximize the achievable data rate and meet the sensing requirements, the authors of \cite{mm_ISAC_1,mm_ISAC_4} have extensively studied the HBF architecture for a mmWave MIMO ISAC system. Authors therein formulated the optimization problem for the hybrid beamformers as the minimization of a weighted sum of the sensing and communication beamforming errors. To solve this problem, Liu \textit{et al.} \cite{mm_ISAC_1} and Yu \textit{et al.} \cite{mm_ISAC_4} proposed the triple alternating approach and the Riemannian optimization methods, respectively, to deal with the constant modulus constraints on the phase-shifters of the analog beamformer. Moreover, the authors of \cite{mm_ISAC_3} optimized the HBF, while considering the required signal-to-interference and noise ratios (SINRs) as quality of service (QoS) constraints of the users. 

The aforementioned works on ISAC-aided mmWave MIMO systems \cite{mm_ISAC_1,mm_ISAC_2,mm_ISAC_4,mm_ISAC_3,mm_ISAC_5} primarily focused on enhancing the sensing and data rate requirements. However, these requirements can not be fulfilled by ignoring the transmit power and energy efficiency (EE). Motivated by these facts, the authors of \cite{co_1,EE_4,EE_1,EE_2,EE_3} have explored the EE of wireless systems. Specifically, Zou \textit{et al.} \cite{EE_1} have considered the problem of EE maximization to optimize the transmit beamformer at the ISAC BS, while constraining the Cramér rao bound (CRB) for the estimation accuracy of the targets. He \textit{et al.} \cite{EE_2} proposed a transmit beamformer design procedure in ISAC systems that maximize the EE of the system under constraints on the SINR of the users and beampattern gains of the targets. Furthermore, the authors of \cite{EE_3} have explored the EE maximization problem in ISAC systems considering the availability of only imperfect channel state information (CSI) of the mobile objects.   

It is important to note that all the above works toward EE optimization of an ISAC system have considered the sub-6 GHz band, and none of them have explored the same for the high-frequency mmWave band. Moreover, transmit beamforming in the above-mentioned works requires a dedicated RFC per antenna, which is highly inefficient for ISAC-enabled mmWave MIMO systems due to a large number of antennas. To deal with this problem, we propose a novel HBF design for ISAC-enabled mmWave MIMO systems to maximize the EE of the system. Particularly, an EE maximization problem is formulated, considering the desired SINR values as the QoS thresholds of the users and beampattern gains for the targets as constraints. To solve this non-convex problem, we first adopt Dinkelbach’s method to obtain the fully-digital beamformer via semi-definite relaxation. Given the fully-digital beamformer, a penalty-based manifold optimization algorithm is proposed in conjunction with the alternating minimization method to optimize the energy-efficient baseband (BB) and analog beamformers.

This paper employs the following notation. Quantities $\mathbf{A}$,  $\mathbf{a}$, and $a$ represent a matrix, a vector, and a scalar quantity respectively;
The $(i,j)$th element, and $i$th element of a matrix $\mathbf{A}$ and a vector $\mathbf{a}$ are denoted by $\mathbf{A}{(i,j)}$ and $\mathbf{a}(i)$, respectively. The conjugate transpose of a matrix $\mathbf{A}$ is denoted by $\mathbf{A}^H$; $\left\vert\left\vert \mathbf{A} \right\vert\right\vert_F$ and $\left\vert a \right\vert$ denote the the Frobenius norm of a matrix and magnitude of a scalar, respectively; $\mathrm{Tr}(\mathbf{A})$ denotes the trace of a matrix $\mathbf{A}$;
$\nabla f$ denotes the gradient vector of function $f$; ${\mathbf I}_M$ denotes an $M \times M$ identity matrix; $\mathcal{CN}(0, \sigma^2)$ is the distribution of a complex circularly symmetric Gaussian random variable with mean $0$ and variance $\sigma^2$.

\section{System Model}\label{mmWave MU MIMO CR System}
\begin{figure}[t]
%\centering
\vspace{-17mm}
\hspace{-6.5mm}
\includegraphics [width=10.1cm]{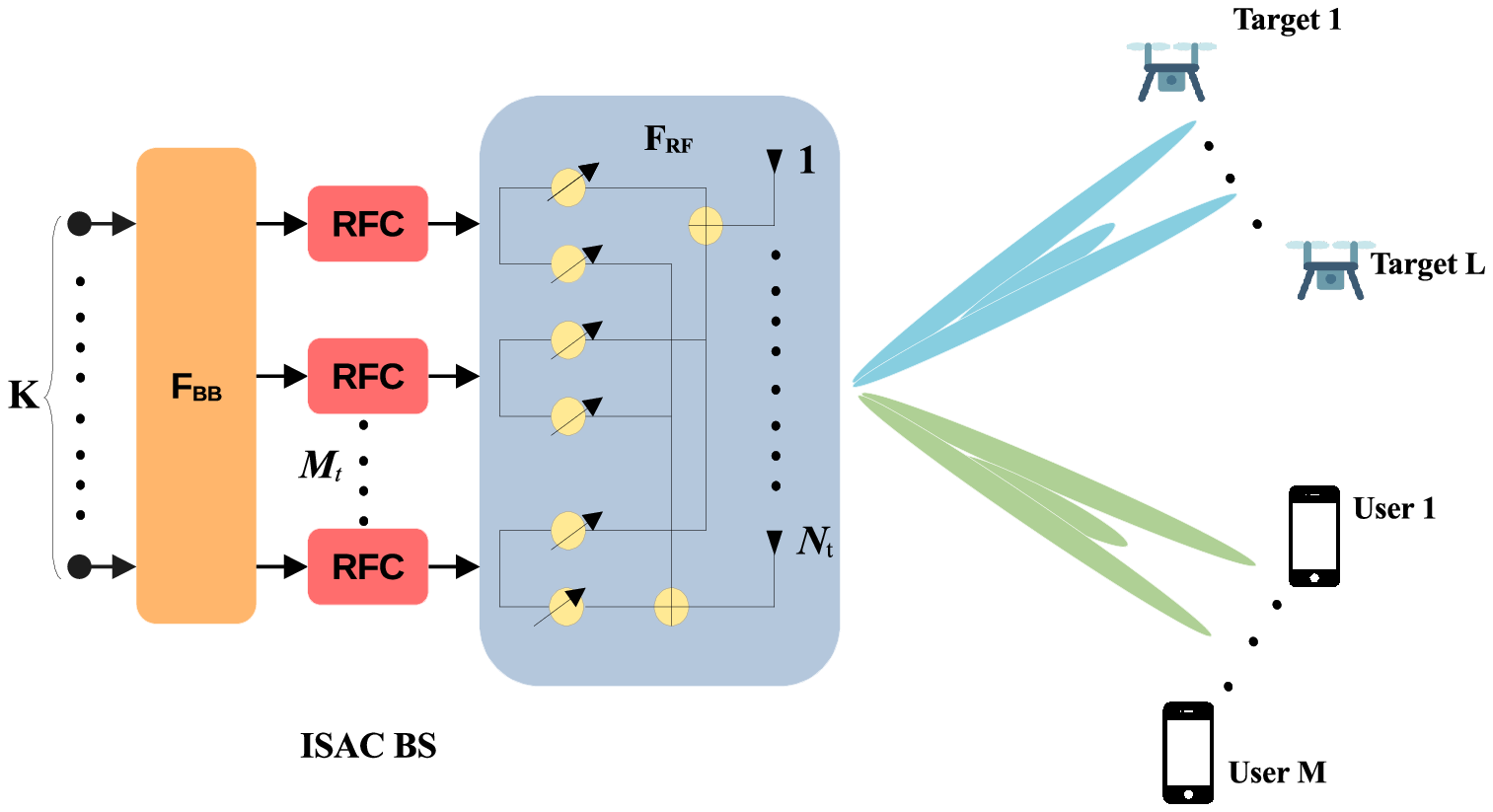}
\vspace{-1.7cm}
\caption{Illustration of an ISAC enabled mmWave MIMO system.}
\label{figure:Fig1}
\vspace{-5mm}
\end{figure}
As shown in Fig. \ref{figure:Fig1}, we consider an ISAC-aided mmWave MIMO system, where
the ISAC BS with $N_\mathrm{t}$ transmit antennas/ receive antennas transmits $K=L+M$ streams to serve $M$ users and detect $L$ different radar targets, simultaneously. Moreover, each user is assumed to be equipped with a single antenna. A fully connected hybrid architecture is assumed to be exploited at the ISAC BS with only $M_\mathrm{t}<< N_\mathrm{t}$ RFCs to reduce the cost and power consumption \cite{mmWave_MIMO_RIS}. Notably, for the considered model, $M_\mathrm{t}$ and $K$ are required to satisfy the property $M_\mathrm{t}\triangleq K$ at the ISAC BS in order to form $L$ beams for the targets and $M$ beams towards the users. 
Let us define the transmit signal $\mathbf{x}\in \mathbb{C}^{K \times 1}$ as 
\begin{equation}
    \mathbf{x} = \begin{bmatrix}\mathbf{s}_1 \\ \mathbf{s}_2\end{bmatrix},
\end{equation}
where $\mathbf{s}_1=[s_1, s_2,\hdots, s_{M}]^T \in \mathbb{C}^{M\times 1}$ is meant for the users and $\mathbf{s}_2=[s_{M+1},\hdots, s_{M_K}]^T\in {\mathbb C}^{L \times 1}$ is used to  detect the targets. Furthermore, we assume that both the signals $\mathbf{s}_1$ and $\mathbf{s}_2$ are statistically independent with zero mean, i.e., satisfying $\mathbb{E}\{\mathbf{x}\}=\mathbf{0}$ and $\mathbf{E}\{\mathbf{x}\mathbf{x}^H\}=\mathbf{I}_{K}$.
Following the fully-connected hybrid architecture \cite{mmWave_MIMO_RIS}, the transmitted signal $\mathbf{x}$ is first precoded by a BB beamformer $\mathbf{F}_\mathrm{BB}=[\mathbf{f}_{{\rm BB},1}, \hdots,\mathbf{f}_{{\rm BB},K}]\in {\mathbb C}^{{M_\mathrm{t}} \times K}$, 
%given by
%%\mathbf{F}_\mathrm{BB}=[\mathbf{f}_{{\rm BB},1}, \hdots,\mathbf{f}_{{\rm BB},K}],
%\end{equation}
followed by precoding using an analog beamformer $\mathbf{F}_\mathrm{RF}\in {\mathbb C}^{{N_\mathrm{t}} \times {M_\mathrm{t}}}$. 
%Moreover, this paper focuses on the energy-efficient hybrid beamforming design, considering only one-time instance while skipping the multiple time instance in the time domain.
%\begin{equation}
%\mathbf{F}_\mathrm{RF}=[\mathbf{f}_{{\rm RF},1}, \hdots,\mathbf{f}_{{\rm RF},M_\mathrm{t}}].
%\end{equation}

\subsection{Communication model}
Considering the availability of the CSI at each user, the received signal $y_m$ at the $m$th user can be written as
\begin{subequations}\label{eqn:rx signal_1}
\begin{align}
&y_m=\mathbf{h}^H_m \mathbf{F}_\mathrm{RF}\mathbf{F}_\mathrm{BB}\mathbf{x} + n_m, \\
=&\mathbf{h}^H_m \mathbf{F}_\mathrm{RF}\mathbf{f}_{\mathrm{BB},m} s_m +
\sum_{n=1, n \neq m}^{K}\hspace{-0.3cm}\mathbf{h}^H_{m}\mathbf{F}_\mathrm{RF}\mathbf{f}_{\mathrm{BB},n} s_n+  n_m,
\end{align}
\end{subequations}
where the quantity $n_m$ is the noise that has the distribution $n_m\sim\mathcal{CN}(0, \sigma^2)$ and $\mathbf{h}_m$ is the narrowband block-fading mmWave MISO channel between the ISAC BS and $m$th user, which is given by the model
\begin{equation}\label{eqn:channel}
\mathbf{h}^H_{m}= \sum_{i=1}^{N^{\rm p}_m}\alpha_{m,i}\mathbf{a}_{\rm t}^H(\theta_{m,i}), 
\end{equation}
where $N^p_m$ denotes the number of multipath components in $\mathbf{h}_m$. The quantity $\alpha_{m,i}$ is the channel gain of the $i$th multipath component with distribution $\mathcal{CN}(0,\beta_m^210^{-0.1PL(d_m)}), \forall l=\{1,\hdots, N^p_m\}$, where $\beta_m=\sqrt{N_\mathrm{t}/N^p_m}$ denotes the normalization factor with $PL(d_m)$ as the path loss that depends on the distance $d_m$ associated with the corresponding link. 
Furthermore, $\mathbf{a}_\mathrm{t}(\theta_l)\in \mathbb{C}^{N_\mathrm{t}\times 1}$ is the steering vector of direction $\theta_l$, which is given by
\begin{equation}\label{eqn:array respo}
\begin{aligned} 
\mathbf{a}_{t}\left (\theta_l \right)=\frac {1}{\sqrt{N_t}}\Biggl [1, \ldots, e^{j \frac {2 \pi }{\lambda } b \left(n \cos \theta_l\right)} ,\ldots, e^{j \frac {2 \pi }{\lambda } b \left ({N_t-1)\cos \theta_l}\right)}\Biggr]^{T},
\end{aligned}
\end{equation}
where, $\lambda$ denotes the wavelength and $b$ represents the antenna spacing, which is assumed to be half of the wavelength.

To simplify the notation, we further define
\begin{equation} \label{eqn:F_1}
\mathbf{F}\triangleq [\mathbf{f}_1, \mathbf{f}_2,\ldots, \mathbf{f}_K] \triangleq \mathbf{F}_{\mathrm{RF}} \mathbf{F}_{\mathrm{BB}}~\in \mathbb {C}^{N_{\mathrm{ t}}\times K},
\end{equation}
where $\mathbf{f}_n \in \mathbb{C}^{N_\mathrm{t}\times 1}, n=1, 2, \hdots, K$ denotes the $n$th column of $\mathbf{F}$.
Thus, the SINR of the $m$th user is given by
\begin{equation}\label{eqn:9}
\gamma_m\left(\mathbf{F}\right) = \frac{\left\vert\mathbf{h}^H_m \mathbf{f}_m\right\vert^2}
{\sum_{n=1, n \neq m}^{K}{\left\vert\mathbf{h}_m^H \mathbf{f}_n\right\vert^2} + \sigma^2 }.
\end{equation}
Based on (\ref{eqn:9}), the achievable rate ${\cal R}_m$ of the $m$th user is given by
\begin{equation}\label{eqn:R_sum}
{\cal R}_m\left(\mathbf{F}\right) = \log_{2}\big(1+\gamma_m\left(\mathbf{F}\right)\big).
\end{equation}

\subsection{Radar model}\label{Radar Model}
We consider the same antenna array to be used at the ISAC BS to transmit and receive radar signals. The resulting signal leakage can be overcome efficiently via the correlation suppression techniques, discussed in \cite{CST}. Thus, the received radar signal $\mathbf {y}_{rad}(\theta_n) \in \mathbb{C}^{N_\mathrm{t} \times 1}$ at the ISAC BS can be written as
\begin{equation} 
\mathbf {y}_{rad}(\theta_n)=\mathbf{r}_{t}+\mathbf{r}_{i}+\mathbf {n}_{r}, 
\end{equation}
where $\mathbf{r}_t, \mathbf{r}_i$ and $\mathbf {n}_{r}$ denote the desired target, interference and the noise signals in the radar sensing environment, respectively. It must be noted that some of the targets may act as scatterers for communication. 
The desired target signal $\mathbf{r}_t$ from the $L$ targets is modeled as
\begin{equation} 
\mathbf{r}_{t}=\sum_{l=1}^{L}\tau_l\mathbf{a}_\mathrm{t}(\theta_l)\mathbf {a}_\mathrm{t}^{T}(\theta _{l})\mathbf{F}\mathbf{s},
\vspace{-2mm}
\end{equation}
where $\tau_l$ is the reflection coefficient for a target located at an angle $\theta_l$. In order to detect multiple targets, the ISAC BS scans different angles of the
space by generating multiple beams toward the targets. To evaluate the sensing performance, we compute the beampattern gains of the targets. Mathematically, the beampattern gain of the target located at $\theta_l$ is given as
\begin{equation}\label{eqn:tx beam pattern}
G(\theta_l, \mathbf{F}) =  \mathbb{E}\left\{\left\vert\mathbf{a}_\mathrm{t}^H(\theta_l)\mathbf{F} \mathbf{x}\right\vert^2\right\} = \mathbf{a}_\mathrm{t}^H(\theta_l) \mathbf{F}\mathbf{F}^H\mathbf{a}_\mathrm{t}(\theta_l).
\end{equation}

\subsection{Energy model}
To evaluate the performance of the communication users, we evaluate the EE of the system in bits/Hz/J, which is defined as the ratio of achievable sum-rate to power consumption. Thus, the EE can be expressed as
\begin{equation}\label{eqn:EE_1}
\begin{aligned}
EE(\mathbf{F}) & =\frac{\sum_{m=1}^M\mathcal{R}_m\left(\mathbf{F}\right)}{P_{\rm diss}\left(\mathbf{F}\right)},
\end{aligned}
\end{equation}
where $P_\mathrm{diss}\left(\mathbf{F}\right)$ is the power dissipation for the considered downlink system. This is given by \cite{EE_1,EE_2}
\begin{equation}\label{eqn:P_diss}
\begin{aligned}
P_{\rm diss}\left(\mathbf{F}\right) = \eta &\sum_{n=1}^K\|\mathbf{f}_n\|^2+ M_\mathrm{t}P_{\rm c},
\end{aligned}
\end{equation}
where $\eta \in [0, 1]$ is the power amplifier efficiency and $P_{\rm c}$ denotes the static hardware power required for each RFC.
\subsection{Problem formulation}
In this work, we aim to optimize the hybrid beamformers $\mathbf{F}_\mathrm{RF}$ and $\mathbf{F}_\mathrm{BB}$ at the ISAC BS, which maximize the EE of the system $EE(\mathbf{F})$. We consider the SINR requirement of each individual user, beampattern gains of the radar targets and total transmit power as constraints. 
Therefore, the pertinent optimization problem is given by
\begin{subequations} \label{eqn:system optimization}
\begin{align}
& \max_{\mathbf{F}}\frac{\sum_{m=1}^M\mathcal{R}_m \left(\mathbf{F}\right)}{P_{\rm diss}\left(\mathbf{F}\right)} \label{eqn:OF_1} \\ 
& \text {s.t.} \quad
\gamma_m (\mathbf{F}) \geq \tau_m, \quad m=1, 2, \hdots, M, \label{constr:SINR}  \\
&\qquad G(\theta_l, \mathbf{F}) \geq \Gamma_l, \quad l=1, 2, \hdots, L,\label{constr:SBP}\\
%&\qquad \left\vert\mathbf{F}_\mathrm{RF}(i,j)\right\vert = 1, \forall i, j, \label{constr:HBF} \\
&\qquad \sum_{n=1}^K\|\mathbf{f}_n\|^2\leq P_\mathrm{t}, \label{constr:TP} 
\end{align}
\end{subequations}
where $\tau_m$ denotes the required SINR threshold of the $m$th user, $\Gamma_l$ is the beampattern gain threshold for the successful sensing of the $l$th target and $P_\mathrm{t}$ is the maximum transmit power at the ISAC BS.
The above problem (\ref{eqn:system optimization}) is highly non-convex, particularly due to the non-concave nature of the objective function (\ref{eqn:OF_1}) and the non-convex constraints (\ref{constr:SINR}). 
\section{Proposed Solution}\label{blind MMSE}
We employ the well-known Dinkelbach’s method \cite{Dink_1} to deal with the non-concavity of (\ref{eqn:OF_1}), which converts a fractional objective function into a subtractive form. 
To this end, we introduce the quantity $\lambda^{\star}$ as the optimal price corresponding to the optimal fully-digital beamformer $\mathbf{F}^{\star}$. Therefore, the objective function (\ref{eqn:OF_1}) can be written in terms of $\lambda^{\star}$ as 
\begin{equation} \label{eqn:q_1}
\lambda^{\star}=\frac{\sum_{m=1}^M\mathcal{R}_m(\mathbf{F}^{\star})}{P_{\rm diss}(\mathbf{F}^{\star})}=\max_{\mathbf{F}}\frac{\sum_{m=1}^M\mathcal{R}_m(\mathbf{F})}{P_{\rm diss}(\mathbf{F})}.
\end{equation}
As a result, the original fractional problem (\ref{eqn:system optimization}) can be reformulated in the subtractive form
\begin{subequations}\label{eqn:EE_3}
\begin{align}
&\max_{\mathbf{F}}\sum_{m=1}^M\mathcal{R}_m(\mathbf{F})-\lambda P_{\rm diss}(\mathbf{F})\\ 
&\text{s.t.} \quad (\ref{constr:SINR}), \hspace{2pt} (\ref{constr:SBP}) \hspace{2pt} \text{and} \hspace{2pt}(\ref{constr:TP}),
\end{align}
\end{subequations}
where $\lambda>0$ regulates the performance of the system between the achievable sum-rate and the EE. When $\lambda=0$, (\ref{eqn:EE_3}) reduces to sum-rate maximization, since the price associated with power dissipation $P_{\rm diss}$ is zero. Whereas, increasing the value of $\lambda$ results in selecting the available power resources wisely to maximize the EE of the system. 

The optimization problem (\ref{eqn:EE_3}) is still non-convex due to the non-convex quantity $\mathcal{R}_m(\mathbf{F})$ and the non-convex constraints in (\ref{constr:SINR}). Therefore, it is challenging to find a closed-form solution using conventional methods.
In order to overcome this hurdle, we follow the MMSE-based approach described in \cite{Dink_2}. Employing the available CSI at the ISAC BS, the quantity $\gamma_m\left(\mathbf{F}\right)$ can be rewritten as
\begin{equation}\label{eqn:9_1}
\gamma_m\left(\mathbf{F}\right) = \mathbf{f}^H_m\mathbf{h}_m \boldsymbol{\Phi}^{-1}\mathbf{h}^H_m \mathbf{f}_m,
\end{equation}
where $\boldsymbol{\Phi} = \sum_{n=1, n \neq m}^K \mathbf{f}_n \mathbf{h}^H_m\mathbf{h}_m \mathbf{f}^H_n + \sigma^2 \mathbf{I}$. Thus, the SINR $\gamma_m\left(\mathbf{F}\right)$ in (\ref{eqn:9}) can be expressed as $\gamma_m(\mathbf{F}) =\mathbf{f}^H_m\mathbf{Q}\mathbf{f}_m, \forall m$, where $\mathbf{Q} = \mathbf{h}_m \boldsymbol{\Phi}^{-1}\mathbf{h}^H_m$.
Hence, the problem (\ref{eqn:EE_3}) can be recast as
\begin{subequations}\label{eqn:EE_4}
\begin{align}
&\max_{\mathbf{F}} \sum_{m=1}^M\log_{2}\left(1+\mathbf{f}^H_m\mathbf{Q}\mathbf{f}_m\right)-\lambda\sum_{n=1}^K\|\mathbf{f}_n\|^2\\ 
&\text{s.t.} \quad \mathbf{f}^H_m\mathbf{Q}\mathbf{f}_m \geq \tau_m, \forall m, \hspace{2pt} (\ref{constr:SBP}) \hspace{2pt} \text{and}\hspace{2pt} (\ref{constr:TP}).
\end{align}
\end{subequations}
Furthermore, we define the Hermitian positive semidefinite matrix $\mathbf{T}_m=\mathbf{f}_m\mathbf{f}^H_m$ that has rank $1$. Therefore, (\ref{eqn:EE_4}) can be reformulated as
\begin{subequations}\label{eqn:EE_5}
\begin{align} 
&\max_{\mathbf{T}_m} \sum_{m=1}^M\log_{2}\left(1+\mathrm{Tr}\left(\mathbf{Q}\mathbf{T}_m\right)\right)-\lambda\sum_{n=1}^K\mathrm{Tr}\left(\mathbf{T}_n\right) \\ 
&\text{s.t.} \quad\quad \mathrm{Tr}\left(\mathbf{Q}\mathbf{T}_m\right)\geq \tau_m, \forall m, \hspace{2pt} (\ref{constr:SBP}) \hspace{2pt} \text{and}\hspace{2pt} (\ref{constr:TP}).
\end{align}
\end{subequations}
We now relax the rank one constraint of $\mathbf{T}_m$ to obtain the semidefinite program (SDP) below
\begin{subequations}\label{eqn:EE_6}
\begin{align}
\max_{\mathbf{T}_m\succeq {\bf 0}} &\sum_{m=1}^M\log_{2}\left(1+\mathrm{Tr}\left(\mathbf{Q}\mathbf{T}_m\right)\right)-\lambda\sum_{n=1}^K\mathrm{Tr}\left(\mathbf{T}_m\right)\\
& {\rm s.t.} \qquad \mathrm{Tr}\left(\mathbf{Q}\mathbf{T}_m\right)\geq \tau_m, \forall m, \hspace{2pt} (\ref{constr:SBP}) \hspace{2pt} \text{and}\hspace{2pt} (\ref{constr:TP}).
\end{align} 
\end{subequations}
Note that the above SDP is convex and can be efficiently solved within polynomial time via standard interior-point methods \cite{Dink_2}. Moreover, it should be noted that the solution provided by the SDP (\ref{eqn:EE_6}) is a sub-optimal solution of the original optimization problem (\ref{eqn:EE_5}). 
However, one can find the rank-one solution by employing eigenvalue decomposition and then choosing the eigenvector corresponding to the maximum eigenvalue as the $m$th beamformer \cite{Dink_2}. 
Next, we obtain the optimal value of the price factor $\lambda$ by employing a local maximizer of problem (\ref{eqn:EE_3}) as 
\begin{equation} \label{eqn:q_3}
\begin{aligned}
\lambda^{\star}=\frac{\sum_{m=1}^M\mathcal{R}_m(\mathbf{F}^{\star})}{P_{\rm diss}.(\mathbf{F}^{\star})}.
\end{aligned}
\end{equation}
\begin{algorithm}[t]
\caption{Energy-efficient hybrid beamformer design for an ISAC mmWave MIMO system}
\label{alg:algo_1}
\begin{algorithmic}[1]
\Require $\mathbf{h}_m, \tau_m, \forall m, \Gamma_l, \forall l$ and desired accuracy $\epsilon_1 \geq 0, \epsilon_2>0, \epsilon_3 > 0$ 
    \State Initialize $\mathbf{F}^{(1)}, \lambda^{(1)}$ and $n=1$ 
    
    \While {$\epsilon \leq \epsilon_1$}
            \State $n \leftarrow n+1$
            \State $\lambda^{(n)} = \frac{\sum_{m=1}^M\mathcal{R}_m(\mathbf{F}^{(n-1)})}{P_{\rm diss}(\mathbf{F}^{(n-1)})}$
            \State Update $\mathbf{F}^{(n)}$ by solving (\ref{eqn:EE_6})
            \State Evaluate $\lambda^{(n)}$ using (\ref{eqn:q_3})
            \State Find $\epsilon = \frac{[\lambda^{(n)}-\lambda^{(n-1)}]}{\lambda^{(n)}}$
    \EndWhile 
    \State \textbf{end while}
    \State Initialize $\mathbf{F}_\mathrm{RF}$ satisfying (\ref{constr:HBF}) and $\mu$
    \Repeat
    \While {$\big \| \widetilde{\mathbf{F}} -\mathbf{F}_{\mathrm{ RF}}\mathbf{F}_{\mathrm{ BB}} \big \|_{\mathrm{ F}}\leq \epsilon_2$}
    \State Obtain $\mathbf{F}_{\mathrm{ RF}}$ via (\ref{EPMO})
    \State Compute $\mathbf{F}_{\mathrm{BB}}$ using (\ref{eqn:LS})
    \EndWhile
    \State \textbf{end while}
    \State Update $\mu=\frac{\mu}{e}, 0<e<1$
    \Until $\left(\big\|\mathbf {F}_\mathrm{RF}\mathbf {F}_\mathrm{BB}\big\|_\mathrm{F}^2-P_\mathrm{t}\right)\leq \epsilon_3$
    \State {\bf return} $\mathbf{F}_\mathrm{RF}, \mathbf{F}_\mathrm{BB}$
\end{algorithmic}
\end{algorithm}

Let the optimal beamformer obtained via the above procedure be denoted as
\begin{equation}    
\widetilde{\mathbf{F}}\triangleq [\widetilde{\mathbf{f}}_1, \widetilde{\mathbf{f}}_2,\hdots, \widetilde{\mathbf{f}}_M]\in \mathbb {C}^{N_{\mathrm{t}}\times M_{\mathrm{t}}}.
\end{equation} 
Consequently, we design the hybrid beamformers $\mathbf{F}_\mathrm{RF}$ and $\mathbf{F}_\mathrm{BB}$ based on $\widetilde{\mathbf{F}}$. One must note that the elements of $\mathbf{F}_\mathrm{RF}$ are constrained to have a constant modulus since these are implemented using phase shifters, as shown in Fig. \ref{figure:Fig1}. As a result, for the given $\widetilde{\mathbf{F}}$, the HBF design problem can be formulated as 
\begin{subequations}\label{eqn:OP_5}
\begin{align}
&\min_{\mathbf{F}_{\mathrm{ RF}},\mathbf{F}_{\mathrm{ BB}}}~\big \| \widetilde{\mathbf{F}} -\mathbf{F}_{\mathrm{ RF}}\mathbf{F}_{\mathrm{ BB}} \big \|^2_{\mathrm{ F}} \label{ob:TPC_1}\\
&~~\mathrm {s.t.}~\big \| \mathbf{F}_{\mathrm{ RF}}\mathbf{F}_{\mathrm{ BB}} \big \|_{\mathrm{ F}}^{2} \leq P_{\mathrm{t}} \label{constr:TP_5}\\
& ~~~~~~~\left\vert\mathbf{F}_\mathrm{RF}(i,j)\right\vert = 1, \nonumber \\ 
&~~~~~~~ i= 1,2,\ldots,N_{\mathrm{t}}, j = 1, 2,\ldots, M_{\mathrm{t}}\label{constr:HBF},
\end{align}
\end{subequations}
where (\ref{constr:HBF}) is the constant modulus constraint on each element of $\mathbf{F}_\mathrm{RF}$. Since $\mathbf{F}_{\mathrm{ RF}}$ and $\mathbf{F}_\mathrm{BB}$ are coupled in the objective function (\ref{ob:TPC_1}) and the constraint (\ref{constr:TP_5}). Hence, we adopt the alternating design approach to solve (\ref{eqn:OP_5}), which is discussed next.
\subsection{Optimization with respect to $\mathbf{F}_\mathrm{RF}$}
For the fixed $\mathbf{F}_\mathrm{BB}$, the optimization problem for $\mathbf{F}_\mathrm{RF}$ can be formulated as follows
\begin{equation}\label{eqn:OP_6}
\begin{aligned}
&\min_{\mathbf{F}_\mathrm{RF}} \big \| \widetilde{\mathbf{F}} -\mathbf{F}_{\mathrm{ RF}}\mathbf{F}_{\mathrm{ BB}} \big \|^2_{\mathrm{ F}} \\
&~~\mathrm {s.t.}~(\ref{constr:TP_5}), (\ref{constr:HBF}).
\end{aligned}
\end{equation}
%To begin with, let us temporarily ignore the power constraint (\ref{constr:TP_5}) while solving (\ref{eqn:OP_5}). Later, we normalize the BB beamformer $\mathbf{F}_\mathrm{BB}$ to satisfy (\ref{constr:TP_5}). Consequently, the optimization problem for design of the analog beamformer $\mathbf{F}_\mathrm{RF}$ for a given $\mathbf{F}_\mathrm{BB}$ is formulated as
To solve the above problem (\ref{eqn:OP_6}), we propose a penalty-based manifold optimization algorithm in which the power constraint (\ref{constr:TP_5}) is relaxed by adding it to the objective function as a penalty term. Subsequently, the problem is solved by the principle of manifold optimization. Thereby, the problem (\ref{eqn:OP_6}) can be converted to the following penalized problem
\begin{equation}\label{EPMO}
\begin{aligned}
& \min \limits_{\mathbf{F}_\mathrm{RF}}~  f(\mathbf {F}_\mathrm{RF}) = \big \| \widetilde{\mathbf{F}} -\mathbf{F}_{\mathrm{ RF}}\mathbf{F}_{\mathrm{ BB}} \big \|^2_{\mathrm{F}} + \mu \left(\big\|\mathbf {F}_\mathrm{RF}\mathbf {F}_\mathrm{BB}\big\|_\mathrm{F}^2-P_\mathrm{t}\right)\\ 
&\mathrm{s. t.} \quad (\ref{constr:HBF}),
\end{aligned}
\end{equation}
where $\mu > 1$ is a penalty factor. Specifically, $\mu$ is obtained via the sequential optimization by increasing the penalty parameter $\mathbf{\mu}$ and solving the problem (\ref{EPMO}) until the solutions eventually converge to the solution of the original problem (\ref{eqn:OP_6}).
Observe that the constraint (\ref{constr:HBF}) represents a Riemannian manifold. Thus, we adopt the Riemannian conjugate gradient (RCG) algorithm \cite{mm_ISAC_4} to solve (\ref{EPMO}), which takes advantage of the Riemannian gradient to evaluate the descent direction.
Toward this end, the Euclidean gradient of the function $f(\mathbf{F}_\mathrm{RF})$ is formulated as
\begin{equation}\label{RCG_3}  
\nabla f(\mathbf{F}_\mathrm{RF})=2\left(\mathbf{F}_\mathrm{RF}\mathbf{F}_\mathrm{BB}\mathbf{F}^H_\mathrm{BB} \left(\mu -1\right) -\widetilde{\mathbf{F}}\mathbf{F}^H_\mathrm{BB}\right).
\end{equation} 
Consequently, one can obtain the Riemannian gradient from the corresponding Euclidean gradient $\nabla f(\mathbf{F}_\mathrm{RF})$, and (\ref{EPMO}) can be solved iteratively on the Riemannian space utilizing the conjugate gradient algorithm. For more details on the RCG algorithm, readers are encouraged to read the paper \cite{mm_ISAC_4}.
\subsection{Optimization with respect to $\mathbf{F}_\mathrm{BB}$}
For a given $\mathbf{F}_\mathrm{RF}$, the optimization with respect to $\mathbf{F}_\mathrm{BB}$ in (\ref{eqn:OP_5}) is given by
\begin{equation}\label{eqn:OP_7}
\begin{aligned}
&\min_{\mathbf{F}_\mathrm{BB}}~\big \| \widetilde{\mathbf{F}} -\mathbf{F}_{\mathrm{ RF}}\mathbf{F}_{\mathrm{ BB}} \big \|_{\mathrm{ F}}.
\end{aligned}
\end{equation}
The solution to the above problem can be obtained employing the well-known least squares (LS) estimate as follows:
\begin{equation} \label{eqn:LS}
\mathbf{F}_{\mathrm{BB}} = (\mathbf{F}_{\mathrm{ RF}}^H \mathbf{F}_{\mathrm{ RF}})^{-1} \mathbf{F}_{\mathrm{RF}}^H \widetilde{\mathbf{F}}.
\end{equation}
We determine $\mathbf{F}_\mathrm{RF}$ and $\mathbf{F}_\mathrm{BB}$ alternatively until convergence is achieved. 
All the above steps in the proposed energy-efficient HBF design for an ISAC-enabled mmWave MIMO system are summarized in Algorithm \ref{alg:algo_1}.

\section{Convergence and complexity analysis}
This section provides a brief analysis of the convergence of Algorithm \ref{alg:algo_1}. One must observe that the convergence of the iterative procedure discussed in Algorithm \ref{alg:algo_1} depends on the price factor $\lambda$.
To show the convergence of $\lambda$, let us assume that $\mathbf{F}^{\star{(n)}}$ is the optimal solution obtained by solving (\ref{eqn:EE_4}) in the $n$th iteration. Hence, one can write
\begin{equation}\label{eqn:q_4}
\begin{aligned} 
& \mathcal{F}\big(\lambda^{(n)}\big)=\sum_{m=1}^M\mathcal{R}_m\big(\mathbf{F}^{\star(n+1)}\big)-\lambda^{(n)}P_{\rm diss}\big(\mathbf{F}^{\star(n+1)}\big)\\ 
&\geq \sum_{m=1}^M\mathcal{R}_m\big(\mathbf{F}^{\star(n)}\big)-\lambda^{(n)}P_{\rm diss}\big(\mathbf{F}^{\star(n)}\big)=0.
\end{aligned}
\end{equation}
Furthermore, substituting the optimal value of $\lambda^{\star}$ using (\ref{eqn:q_3}), one can recast (\ref{eqn:q_4}) as
\begin{equation} 
P_{\rm diss}\big(\mathbf{F}^{\star(n+1)}\big)\times\big(\lambda^{(n+1)}-\lambda^{(n)}\big)\geq 0.
\end{equation}
Since the quantity $P_{\rm diss}\big(\mathbf{F}^{\star(n+1)}\big) \geq 0$, it follows that $\lambda^{(n+1)} \geq \lambda^{(n)}$. Therefore, $\lambda$ is a non-decreasing function with an iteration index of $n$, and thus converges after some iterations. Furthermore, step 11 of Algorithm \ref{alg:algo_1} dominates the computational cost of $\mathbf{F}_\mathrm{RF}$, which arises due to evaluation of the Euclidean gradient (\ref{RCG_3}). Therefore, the computational complexity for obtaining the gradient (\ref{RCG_3}) is given by $\mathcal{O}(M_\mathrm{t}N^2_\mathrm{t}K)$.
\section{\uppercase{Simulation Results}}\label{simulation results}
In this section, we present the simulation results to characterize the performance of our proposed energy-efficient HBF design of the ISAC-enabled mmWave MIMO system operating at the carrier frequency of $28$ GHz. 
We consider the users and the targets to be located at $\left[30^\circ, 60^\circ\right]$ and $\left[-60^\circ, -20^\circ\right]$, respectively. The pathloss model $PL(d_m)$ of the mmWave MIMO channel is given by \cite{mmWave_MIMO_RIS}
\begin{equation}\label{eqn:path loss model}
\begin{aligned}
PL(d_m)\hspace{0.02in}[\rm dB] = \varepsilon + 10\varphi\log_{10}(d_m)+\varpi,
\end{aligned}
\end{equation}
where we have $\varpi \in {\cal CN}(0,\sigma_{\rm \varpi}^2)$ with $\sigma_{\rm \varpi}=5.8 \hspace{0.02 in}{\rm dB}$, $\varepsilon=61.4$ and $\varphi=2$ \cite{mmWave_MIMO_RIS}.  
Moreover, the angle of departure $\theta_{m,i}, \forall m, i$ is generated from a truncated Laplacian distribution with uniformly-random mean angle of $\overline{\theta}$ and a constant angular spread of $\frac{\pi}{2N_\mathrm{t}}$.
The noise variance and SINR threshold of each user are set as $\sigma^2=\sigma^2_m$ and $\tau = \tau_m, m=1, \hdots, M$, respectively. In similar fashion, the beampattern gain threshold of each target is set as $\Gamma = \Gamma_l, l = 1, \hdots, L$. Furthermore, the quantity $\mathrm{SNR}$ is defined as $\frac{P_{\rm t}}{\sigma^2}$, where $P_\mathrm{t}$ is the transmit power. The specific values of the simulation parameters are listed in Table \ref{tab:simulation parameters}.
Furthermore, we compare the performance of our proposed HBF design with successive interference cancellation (SIC)-based HBF \cite{co_1}, orthogonal matching pursuit (OMP)-based HBF \cite{co_2}, and the fully-digital beamforming (FDB) schemes.
\begin{table}[t]
  \centering
  \caption{Simulation parameters and corresponding values} 
\label{tab:simulation parameters}
  \begin{tabular}{|P{1.5cm}|P{4cm}|P{1.5cm}|}
  \hline
  Notation & Parameter & Value \\
  \hline
  $N_\mathrm{t}$ & Number of transmit antennas & $64$\\
  \hline
   $M_\mathrm{t}$ & Number of RFCs & $4$\\
  \hline
$M$ & Number of communication users & $2$\\
  \hline
  $L$ & Number of targets & $2$\\
  \hline
  $N_m^{\mathrm p}, \forall m$ & Number of propagation paths & $10$\\
  \hline
  $P_\mathrm{t}$ & Maximum transmit power & $20$ dB\\
  \hline
  $\sigma^2$ & Noise power & $-91$ dBm\\
  \hline
  $\tau$ & SINR threshold & $10$ dB\\
  \hline
  $\Gamma$ & Beampattern gain threshold & $5$ dB\\
  \hline
  $\eta$ & Amplifier efficiency & $0.3$ \cite{EE_1}\\
  \hline
  $P_\mathrm{c}$ & Static hardware power & $30$ dBm \cite{EE_1}\\
  \hline
  $\epsilon_1, \epsilon_2, \epsilon_3$ & Desired accuracy & $10^{-3}$\\
  \hline
 \end{tabular}
 \vspace{-5mm}
 \end{table}
 \begin{figure}[t]
\centering
\includegraphics[width=0.8\columnwidth]{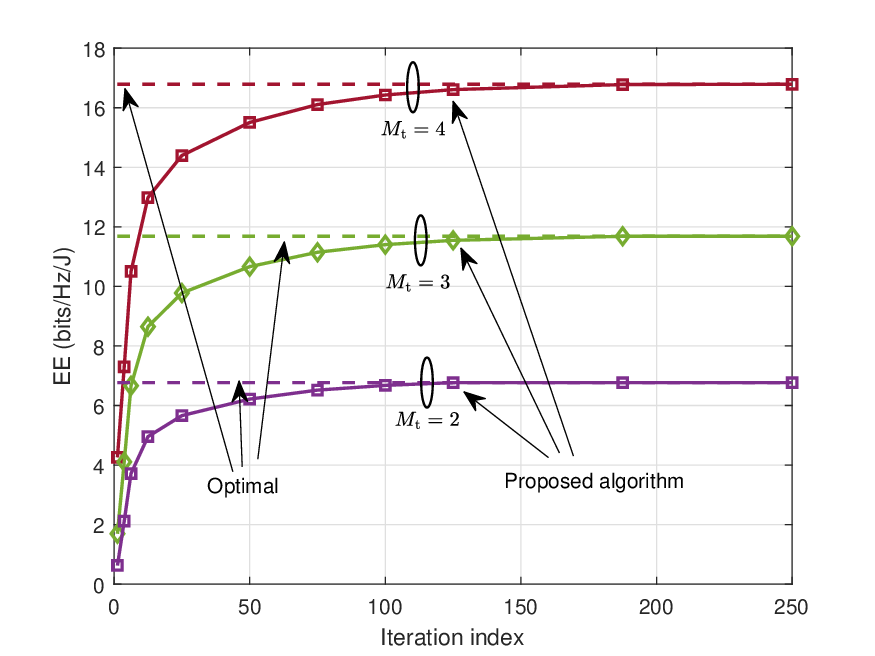}%
\caption{Convergence of the proposed algorithm.}
\label{fig:R1}
\vspace{-7mm}
\end{figure}

In Fig. \ref{fig:R1}, we investigate the convergence behavior of the proposed algorithm by plotting the EE versus the maximum number of iterations for different values of the number of RFCs ($M_\mathrm{t}$). As seen from the figure, the EE converges after reaching its maximum value after a few iterations. It can also be observed that with an increase in $M_\mathrm{t}$, the number of iterations required for convergence to obtain the optimal RF and BB precoders also increases.
\begin{figure*}[t]%
\centering
\begin{subfigure}{.5\columnwidth}
\includegraphics[width=1.1\columnwidth]{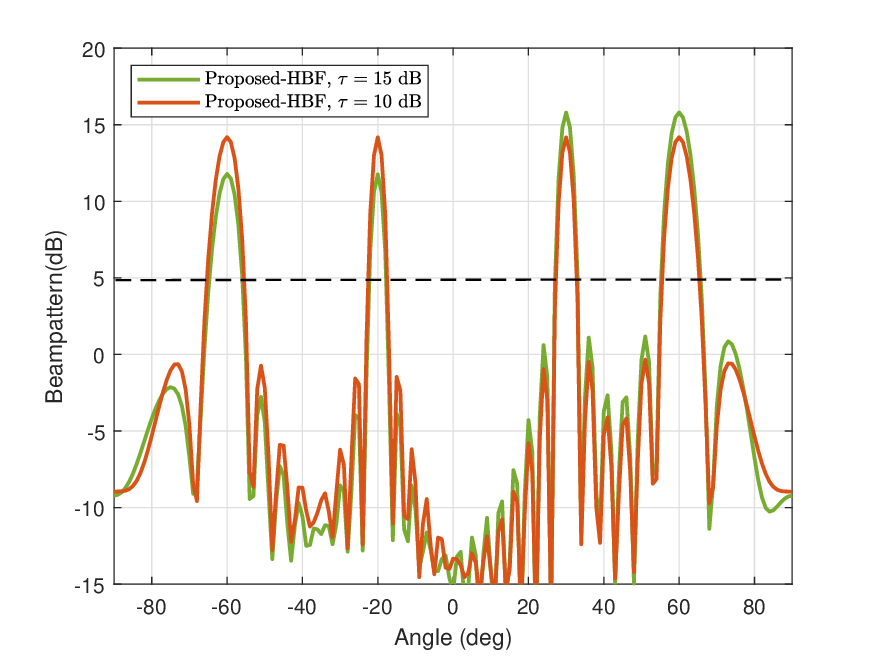}%
\caption{%Sum-SE versus $\mathrm{SNR}$.
}
\label{fig:R2}
\end{subfigure}%\hfill%
\begin{subfigure}{.5\columnwidth}
\includegraphics[width=1.1\columnwidth]{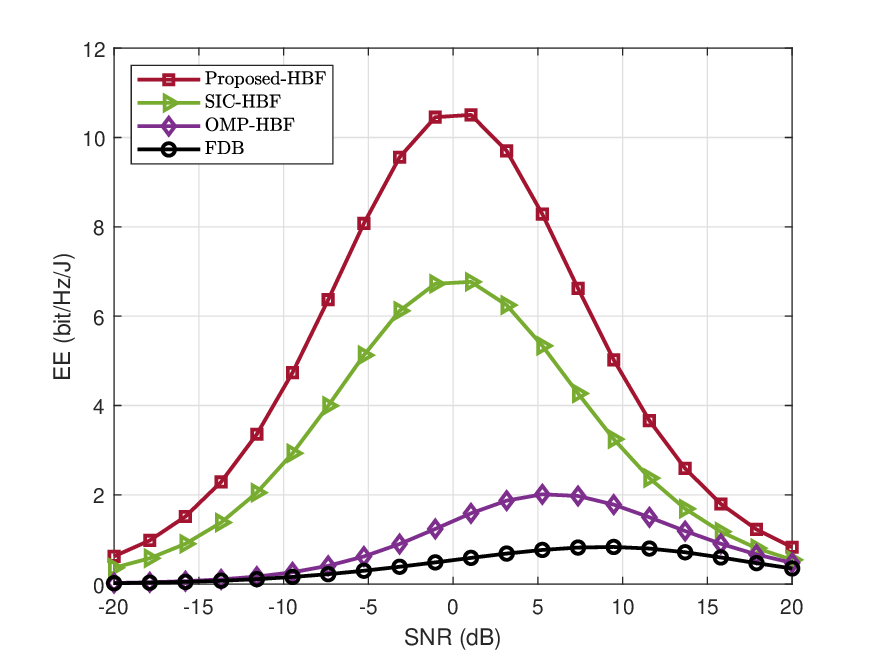}%
\caption{%Sum-SE versus similarity threshold $\epsilon$.
}
\label{fig:R3}
\end{subfigure}%
%\label{fig:hasil}
\begin{subfigure}{.5\columnwidth}
\includegraphics[width=1.1\columnwidth]{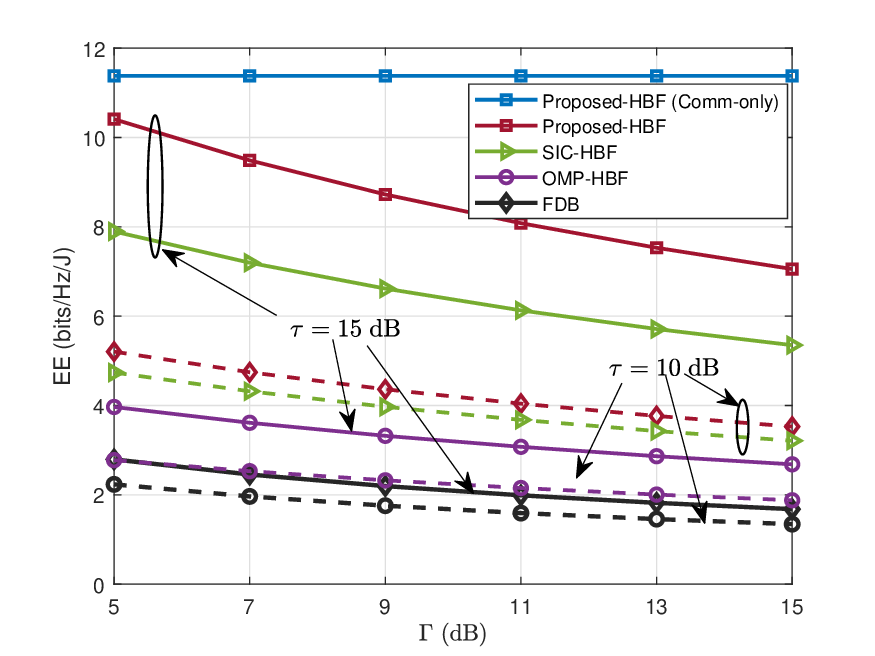}%
\caption{%Sum-SE versus number of TAs $N_\mathrm{t}$.
}
\label{fig:R4}
\end{subfigure}%
\begin{subfigure}{.5\columnwidth}
\includegraphics[width=1.1\columnwidth]{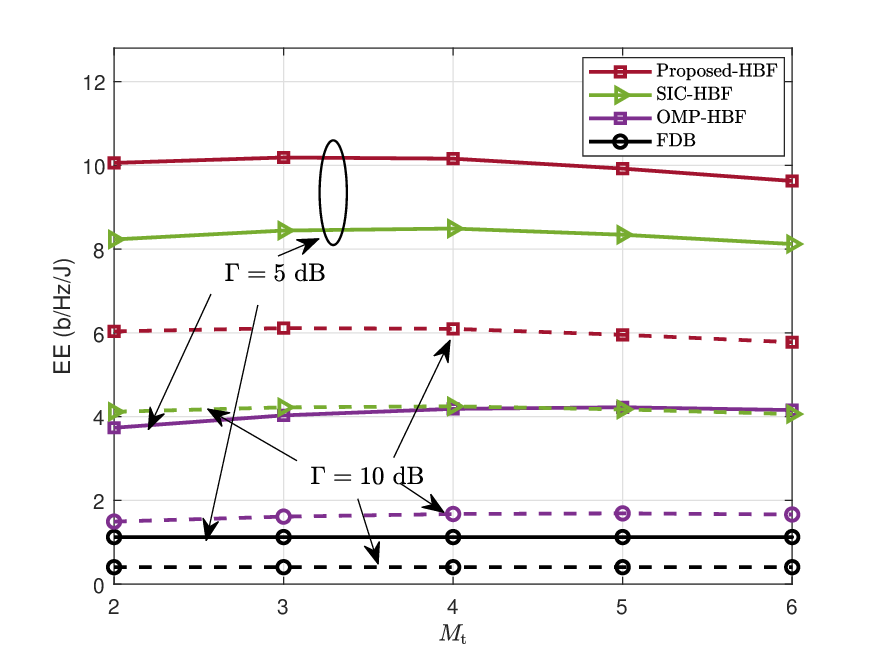}%
\caption{%Sum-SE versus similarity threshold $\epsilon$.
}
\label{fig:R5}
\end{subfigure}%
\caption{(a) Transmit beampattern for different $\mathrm{SINR}$ thresholds $\tau$; (b) EE versus $\mathrm{SNR}$; (c) EE versus beampattern gain $\Gamma$ for different $\mathrm{SINR}$ thresholds $\tau$; (d) EE versus number of RFCs $M_\mathrm{t}$ for different beampattern gain $\Gamma$. }
\vspace{-6mm}
\end{figure*}
%\begin{figure}[t]
%\centering
%\includegraphics[width=\columnwidth]{R1.eps}%
%\caption{Transmit beampattern of the ISAC BS with the proposed HBF for different SINR constraints.}
%\label{fig:R1}
%\vspace{-5mm}
%\end{figure}

In Fig. \ref{fig:R2}, we plot the transmit beampattern of the ISAC BS. It can be seen from the figure that the main lobes of the ISAC BS are focused towards the users as well as the targets. Moreover, as the required SINR threshold of the users increases from $\tau=10$ dB to $\tau=15$ dB, the beampattern gain towards the users increases, which is in line with our expectations. Also, observe that the beampattern gain of the targets is higher than the required threshold of $\Gamma = 5$ dB.
%\begin{figure}
%\centering
%\includegraphics[width=\columnwidth]{R2.eps}%
%\caption{Comparison of the EE of the proposed HBF with different schemes in terms of SNR.}
%\label{fig:R2}
%\vspace{-5mm}
%\end{figure}%

Fig. \ref{fig:R3} shows the EE of the system versus SNR. As can be seen from the figure, the EE increases first with increasing SNR, reaches its optimal value, and then further decreases with increasing SNR. This is due to the fact that with increasing transmit power, the resultant SNR increases, which enhances the achievable rate. However, after a certain transmit power, the energy consumption increases more rapidly than the rate, which results in a decrease in the EE. Furthermore, our proposed HBF design performs better than the benchmarks, which shows the effectiveness of the RCG and LS methods used to approximate the hybrid beamformers $\mathbf{F}_\mathrm{RF}$ and $\mathbf{F}_\mathrm{BB}$ with the optimal beamformer $\mathbf{\widetilde{F}}$. Additionally, the EE of the FDB is worse than the other competing schemes, a natural result of its requirement for a large number of power-hungry RFCs.
%\begin{figure}[t]
  %   \centering
  %   \begin{subfigure}[b]{0.24\textwidth}
   %      \centering
   %      \includegraphics[width=\textwidth]{R_2.eps}
    %     \caption{}
    %     \label{fig:R3}
    % \end{subfigure}
    % \begin{subfigure}[b]{0.24\textwidth}
    %     \centering
    %     \includegraphics[width=\textwidth]{R4.eps}
     %    \caption{}
     %    \label{fig:R4}
    %\end{subfigure}
     %\caption{Comparison of the EE of the proposed HBF with different schemes in terms of (a) beampattern gain constraint $\Gamma$; (b) number of RFCs $M_\mathrm{t}$.}
    % \vspace{-2mm}
%\end{figure}

%\begin{figure}[t]
%\centering
%\includegraphics[width=\columnwidth]{R3.eps}%
%\caption{Comparison of the EE of the proposed HBF with different schemes in terms of beampattern gain constraint.}
%\label{fig:R3}
%\vspace{-5mm}
%\end{figure}

In Fig. \ref{fig:R4}, we depict the EE in terms of the beampattern gain threshold $\Gamma$ and also benchmark it with the proposed-HBF scheme used for only communication purposes, i.e., without the beampattern gain requirements of the targets. As the beampattern gain $\Gamma$ of the targets increases, the EE of the system decreases, as a high value of $\Gamma$ increases the power towards the targets, leading to a decrease in the rate of the users. Furthermore, the EE of the proposed HBF-scheme used only for communication remains constant with respect to $\Gamma$ and acts as a benchmark for the users. Moreover, the EE of the system improves with increasing $\mathrm{SINR}$ threshold from $\tau = 10$ dB to $\tau = 15$ dB, which is in line
with our expectations.
%\begin{figure}
%\centering
%\includegraphics[width=\columnwidth]{R4.eps}%
%\caption{Comparison of the EE of the proposed HBF with different schemes in terms of number of RFCs at the ISAC BS.}
%\label{fig:R4}
%\vspace{-5mm}
%\end{figure}%

Finally, we plot the EE versus number of the RFCs $M_\mathrm{t}$ in Fig. \ref{fig:R5} to study the trade-off between the EE and $M_\mathrm{t}$. The EE of the system first increases with $M_\mathrm{t}$ and then decreases after a certain point. This is due to the fact that an increasing value of $M_\mathrm{t}$ leads to an increase in the achievable-rate, but also increases the energy consumption. Hence, an optimal number of RFCs is required to maximize the EE of the ISAC-enabled mmWave MIMO system for the given power, SINR, and beampattern gain thresholds. Moreover, the proposed scheme yields improved performance over the benchmarks at low as well as at high beampattern gain thresholds $\Gamma=\{5, 10\}$ dB, which demonstrates the efficacy of the penalty-based manifold optimization technique for HBF design.
\section{\uppercase{Conclusion}}\label{conclusion}
In this paper, we explored the EE maximization problem of an ISAC-enabled mmWave MIMO system.
A Dinkelbach’s method-based alternating minimization algorithm was proposed, which first optimized the transmit beam while considering the SINR and beampattern gain thresholds as the constraints. Subsequently, the BB and analog beamformers are designed to minimize the beamforming error between the optimal transmit beam and hybrid beamformers under a constant modulus constraint on each element of the RF beamformer. Finally, simulation results are shown and compared with the benchmarks under different settings, which show the efficacy of our proposed HBF design in terms of EE performance.
\bibliographystyle{IEEEtran}
\bibliography{biblio.bib}
\end{document}